\documentclass[aps, prb, reprint, amssymb, amsfonts]{revtex4-1}
\usepackage{amsmath}
\usepackage{bm}
\usepackage{graphicx}
\usepackage[unicode]{hyperref}
\usepackage[latin1]{inputenc}
\usepackage{array}

\begin{document}

\title{Spin-mediated dissipation and frequency shifts of a cantilever at milliKelvin temperatures}

\author{A. M. J. den Haan}
\author{J. J. T. Wagenaar}
\author{J. M. de Voogd}
\author{G. Koning}
\author{T. H. Oosterkamp}
	\email{oosterkamp@physics.leidenuniv.nl}
\affiliation{Kamerlingh Onnes Laboratory, Leiden University, PO Box 9504, 2300 RA Leiden, The Netherlands} 

\begin{abstract}
We measure the dissipation and frequency shift of a magnetically coupled cantilever in the vicinity of a silicon chip, down to $25$ mK. 
The dissipation and frequency shift originates from the interaction with the unpaired electrons, associated with the dangling bonds in the native oxide layer of the silicon, which form a two dimensional system of electron spins. We approach the sample with a $3.43$ $\mu$m-diameter magnetic particle attached to an ultrasoft cantilever, and measure the frequency shift and quality factor as a function of temperature and the distance. Using a recent theoretical analysis [J. M. de Voogd et al., arXiv:1508.07972 (2015)] of the dynamics of a system consisting of a spin and a magnetic resonator, we are able to fit the data and extract the relaxation time $T_1=0.39\pm0.08$ ms and spin density $\sigma=0.14\pm0.01$ spins per nm$^2$. Our analysis shows that at temperatures $\leq500$ mK magnetic dissipation is an important source of non-contact friction.
\end{abstract}

\maketitle

Understanding the dissipation and frequency shifts in magnetic force experiments is crucial for the development of magnetic imaging techniques, e.g. Magnetic Resonance Force Microscopy (MRFM). The sensitivity of such techniques depends on the friction of the cantilevers, which therefore has increased the interest in high-quality cantilevers with quality factors exceeding a million \cite{Tao2014}. However, the quality factor reduces due to non-contact friction with the scanned sample which is explained by dielectric fluctuations \cite{Kuehn2006}. Far from the surface, magnetic dissipation from paramagnetic spins or nanomagnets on the cantilever have been observed to have a large effect on the friction \cite{Stipe2001, Harris2003}.  Our report quantitatively analyzes the magnetic dissipation of a cantilever in the vicinity of a silicon chip, showing that this is the most significant non-contact friction at low temperatures for a magnet on cantilever geometry.

Magnetic Force Microscopy (MFM) measures the forces resulting from stray fields of a sample that is being scanned. The coupling of the tip with the magnetic field manifests itself as a shift in the resonance frequency of the cantilever and as additional dissipation which reduces its quality factor $Q$. For magnetic moments that do not change due to the magnetic tip itself, the frequency shifts are well understood. However, a more complicated model is required when the spins in the sample are paramagnetic, because the motion of the tip changes the direction of their  magnetic moments \cite{Rugar1990}.

\begin{figure}[bt]
\includegraphics[width=\columnwidth]{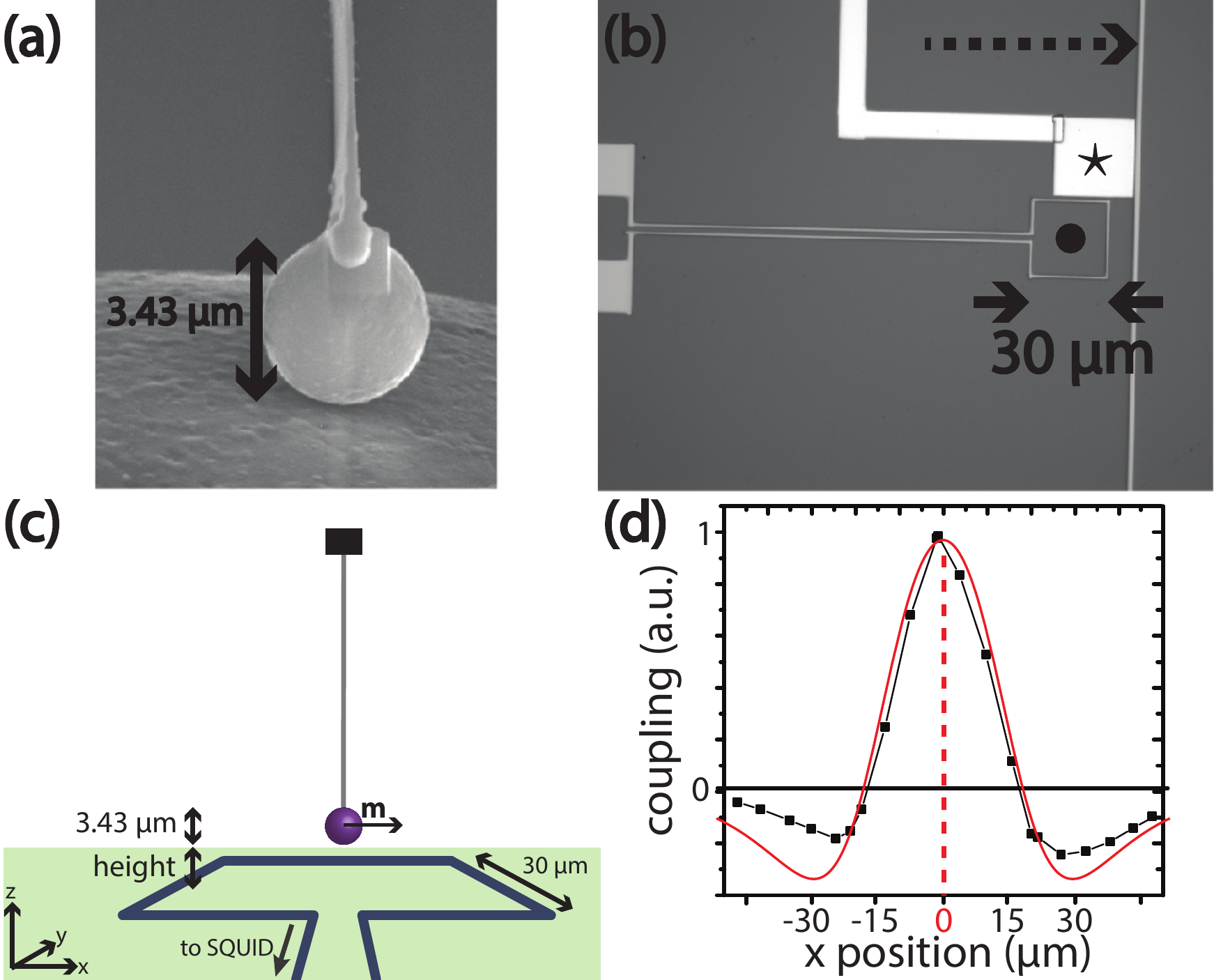}
\caption{ (a) Scanning Electron Microscope image of the magnetic particle after it is glued to the cantilever. (b) Optical microscope image of the detection chip. The cantilever is positioned above the center of the pickup coil ($\bullet$). The pickup coil is used for SQUID based detection of the cantilever's motion. The vertical wire (dotted arrow) and the copper sample ($\star$) are used in other experiments. (c) Sketch of the setup. The height is measured from the bottom of the magnetic particle, which has a diameter of 3.43 $\mu$m. (d) The coupling with the pickup coil as function of the x-position of the cantilever. The red solid line is the calculated flux change in a square loop due to a magnetic dipole $\bm{\mu}$ on a moving resonator. The maximum (scaled to 1) of the curve is at the center of the pick-up coil, which can be determined with $\mu$m precision.}
\label{figure:figure1}
\end{figure}

In this paper, we show frequency shifts and dissipation resulting from the dangling electron bonds at the surface of a silicon substrate.  We are able to extract the relaxation time $T_1$ of the electron spins, without using electron spin resonance techniques. Furthermore, we use our analysis to calculate the maximum possible dissipation for a state-of-the-art MRFM setup and diamond cantilever. We show that magnetic dissipation can cause a drop in quality factor, thereby decreasing the sensitivity of an MRFM experiment. We calculate that this dissipation is suppressed when using large external magnetic fields at low temperatures.

In our experiment, a magnet attached to a cantilever (Fig. \ref{figure:figure1}a) couples via its magnetic field $\bm{B}(\bm{r})$ to magnetic moments $\bm{\mu}$ originating from localized electron spins with near-negligible interactions. The coupling with a single spin can be associated with a stiffness $k_s$, which results in a shift $\Delta f$ of the natural resonance frequency $f_0$ of the cantilever, according to $\Delta f=\frac{1}{2}\frac{k_s}{k_0}f_0$, with $k_0$ the natural stiffness of the cantilever. %Furthermore, it will affect the mechanical quality factor $Q$ of the cantilever.

Commonly, the analysis of magnetic interaction \cite{Rugar2004} begins with the interaction energy $E=-\bm{\mu}\cdot\bm{B}(\bm{r})$. And one calculates the force and stiffness acting on the cantilever by taking the first and second derivative with respect to x, the direction of the fundamental mode of the cantilever. Assuming that $\bm{\mu}$ is fixed by a large external field, one obtains in this approach a stiffness in the form of $k_s=\bm{\mu}\cdot\frac{\partial^2 \bm{B}(\bm{r})}{\partial x^2}$.

\begin{figure}[tb]
\includegraphics[width=\columnwidth]{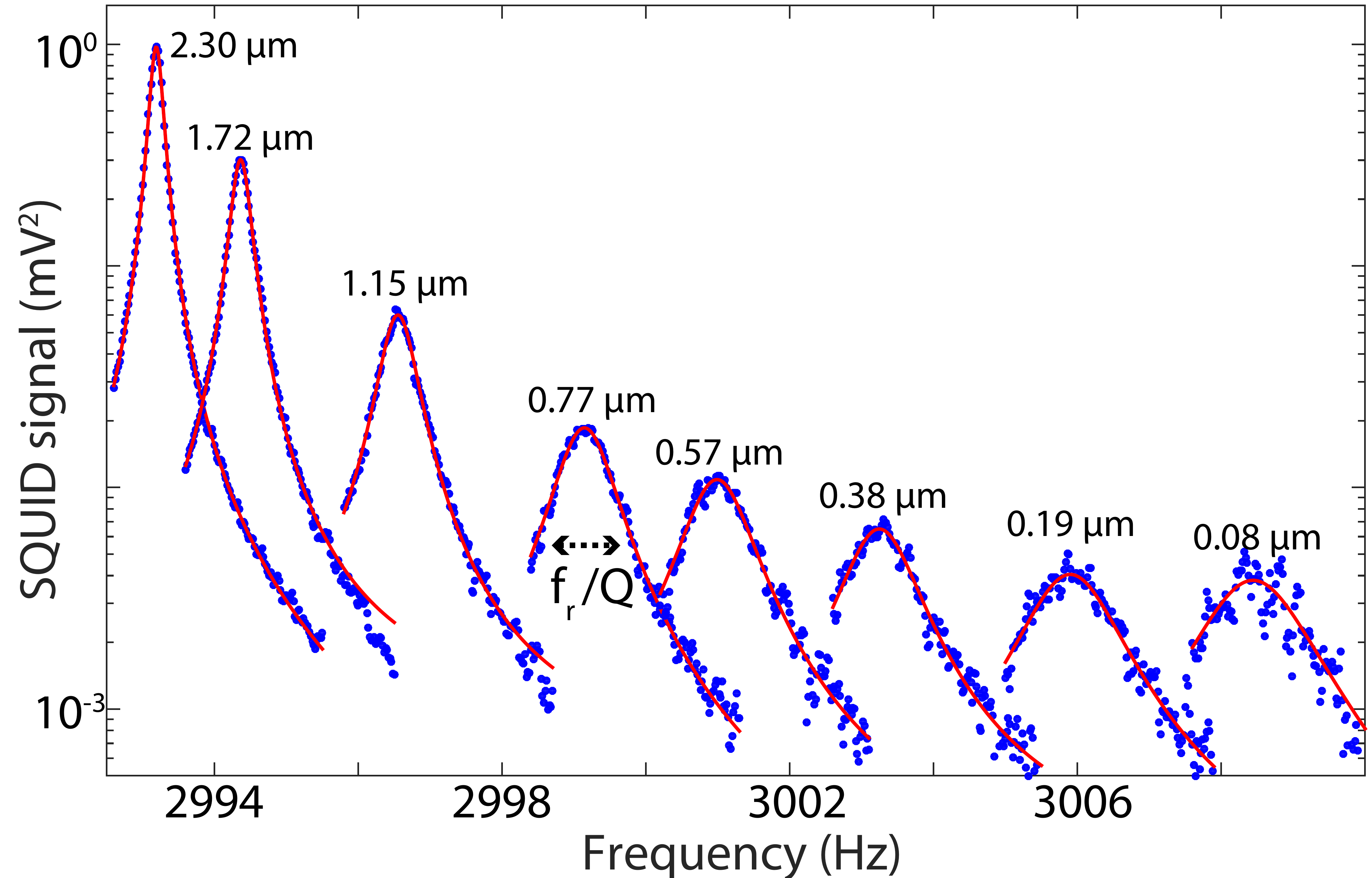}
\caption{Frequency sweeps of the cantilever at a temperature of $70$ mK. When moving towards the sample, the resonance frequency $f_r$ increases, while $Q$ decreases due to an increasing coupling with the surface electron spins. We extract $f_r$ and $Q$ by fitting the data to a Lorentzian (red solid line).}
\label{figure:figure2}
\end{figure}

A recent detailed analysis by De Voogd et al. \cite{DeVoogd2015}, which starts with the Lagrangian of the full system, taking into account the spin's dynamics as well as the influence of the mechanical resonator on the spin, suggests that the commonly employed model is not the correct approach for paramagnetic spins. For paramagnetic spins, the relaxation and the exact dynamics of the spin in the cantilever's magnetic field determine the frequency shifts and dissipation. In the case of a two-dimensional system of paramagnetic spins, uniformly distributed over an infinite surface, the frequency shift $\Delta f$ and shift in the inverse quality factor $\Delta \frac{1}{Q}$ can be written as: 

\begin{equation}
\frac{\Delta f}{f_0}= \frac{1}{2} C\cdot \frac{\left(2\pi f_0T_1\right)^2}{1+\left(2\pi f_0T_1\right)^2},
\label{eq:1}
\end{equation}
\begin{equation}
\Delta \frac{1}{Q}=C\cdot \frac{2\pi f_0T_1}{1+\left(2\pi f_0T_1\right)^2},
\label{eq:2}
\end{equation}
\begin{equation}
C=\frac{\sigma \mu^2}{k_0 k_B T}\iint_S\frac{\left(\hat{\bm{B}}(\bm{r})\cdot \frac{\partial \bm{B}(\bm{r})}{\partial x}\right)^2}{\cosh^2{(\frac{\mu B(\bm{r})}{k_B T})}} d\bm{r}.
\label{eq:3}
\end{equation}

Where $T$ is the temperature, $k_B$ is the Boltzmann constant and $T_1$ is the spin's longitudinal relaxation time. The integral is performed over the infinite surface assuming a constant spin density $\sigma$. We have assumed $\Delta f\ll f_0$, $Q\gg 1$, and that the inverse of the transverse relaxation time $T_2^{-1}$ is much smaller than the Larmor frequency, which is already the case when $T_2$ is larger than $1$ $\mu$s.

In this paper, we detect the dangling bonds that are present on the surface of a silicon substrate of the detection chip using MFM down to $25$ mK. We use a commercial cryogen-free dilution refrigerator, in which we implemented several vibration isolation measures \cite{DenHaan2014}. We are able to coarse approach towards the sample in three dimensions, with a range of $1$ mm in x, y and z. For this we employ three `PiezoKnobs', from Janssen Precision Engineering B.V., while reading out the position using three capacitive sensors. 

The cantilever is a silicon micro-machined IBM-type with length, width and thickness of 145 $\mu$m, $5$ $\mu$m and $100$ nm, respectively \cite{Chui2003,Mamin2003}. The magnetic particle is a spherical particle from a commercial neodymium-alloy powder \footnote{The neodymium-alloy powder is of type MQP-S-11-9-20001-070 by Magnequench, Singapore}. We used platinum electron beam induced deposition using an in-house developed nanomanipulator \cite{Heeres2010} in a Scanning Electron Microscope (SEM) to attach the small magnetic particle on the free end of the cantilever and measured the diameter to be 3.43 $\mu$m (Fig. \ref{figure:figure1}a). Subsequently, we magnetized the magnet in the x-direction at room temperature in a field of $5$ T.

The readout of the cantilever's motion is based on a Superconducting Quantum Interference Device (SQUID) which enables low temperature experiments  \cite{Usenko2011}. Where in conventional MFM setups a laser is used to readout the motion, our method is based on the motion of the magnetic particle in the vicinity of a small superconducting `pickup' coil, giving a flux change whenever the cantilever moves (Fig. \ref{figure:figure1}c). This signal is transformed by an on-chip transformer, which matches the pickup coil inductance to the high SQUID input inductance. The measured flux noise in the complete setup is less than $4$ $\mu \Phi_0/\sqrt{Hz}$, where $\Phi_0$ is the flux quantum.

The substrate is high resistivity ($>1$ k$\Omega$cm) (100)-oriented n-type (phosphorus doped) silicon. The substrate is cleaned with acetone and DI water, which leaves an interface of silicon with its native oxide. To create the superconducting structures on the chip, NbTiN is grown on the silicon substrate with a thickness of roughly $300$ nm. Patterning is done using standard nano-lithographic techniques and reactive ion etching in a SF$_6$/O$_2$ plasma. For future MRFM experiments, we added a wire for radio-frequency currents and a $300$ nm thick copper layer capped with gold. The copper is connected via golden wire bonds to the sample holder, which itself is connected via a silver welded wire to the mixing chamber, ensuring good thermalization of the sample. Figure \ref{figure:figure1}b shows an optical microscope image of the obtained structure.

\begin{figure}[tb]
\includegraphics[width=\columnwidth]{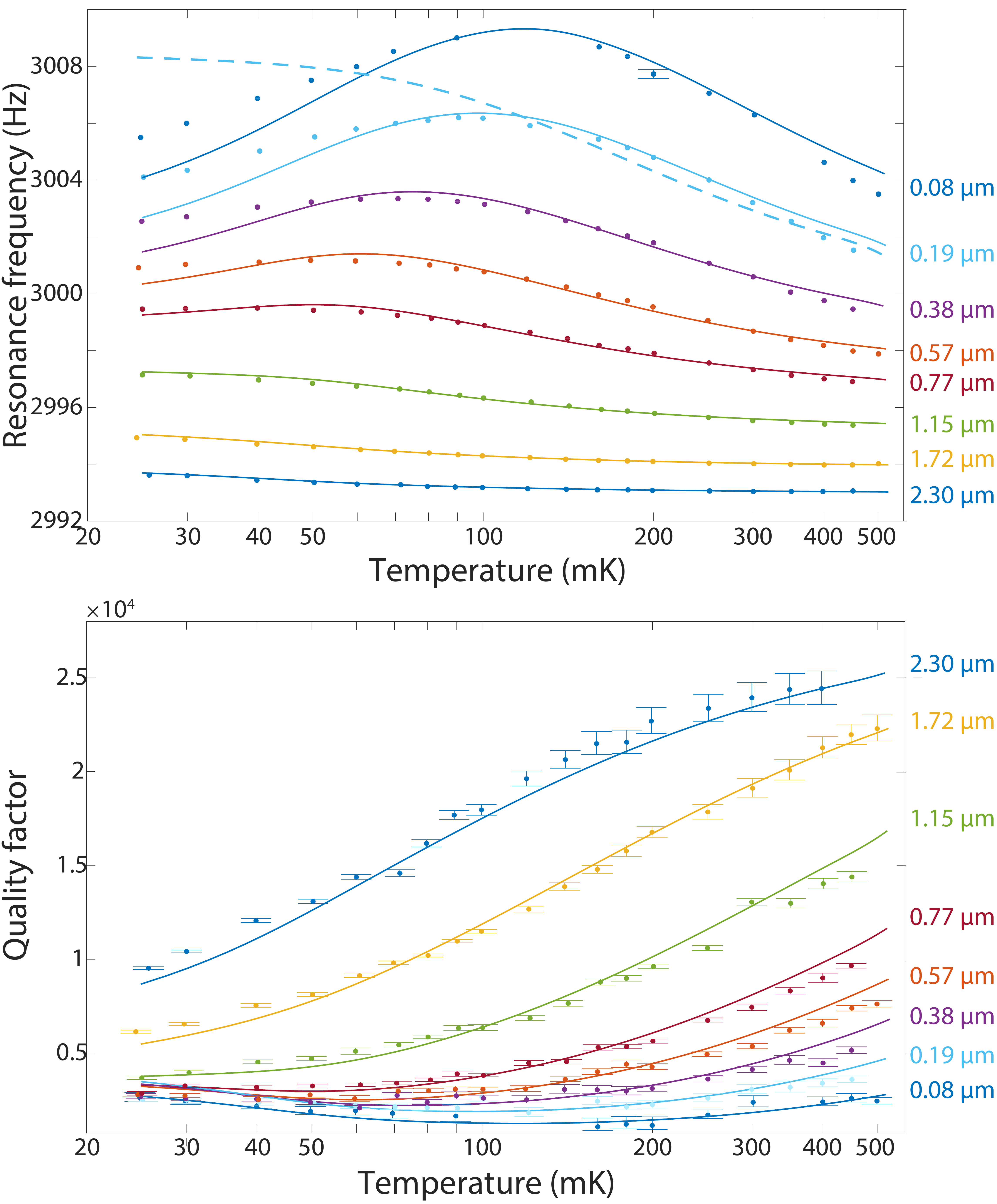}
\caption{Resonance frequency $f_r$ and quality factor $Q$ versus temperature for different heights of the cantilever with respect to the sample. For the quality factor, the error bars indicate the $95$ \% confidence intervals of the Lorentzian fit. For the frequencies the average error was $0.01$ Hz, which is smaller than the point size, except for one data point. The solid lines are fits to the data with the spin density $\sigma$, spin relaxation time $T_1$ and frequency offset $f_0$ as fitting parameters. $f_r$ and $Q$ are simultaneously fitted for each height. The results of the fit can be found in table \ref{tabel}. The dashed line is the frequency shift calculated with the commonly used expression $k_s=p \bm{\mu} \cdot \frac{d^2 \bm{B}}{dx^2}$, with $p=\tanh{\left(\frac{\mu B(\bm{r})}{k_BT}\right)}$ and with $\sigma$ ten times smaller than we find in our analysis.}
\label{figure:figure3}
\end{figure}

We drive the cantilever using a small piezo element glued to the cantilever holder. We sweep the drive frequency using a function generator around the resonance frequency $f_r$ while measuring the SQUID's response using a Lock-In amplifier. We fitted the square of the SQUID's signal with a Lorentzian curve in order to extract $f_r$ and $Q$. The amplitude of the Lorentzian is determined by the coupling between the magnet and the pickup coil, which is proportional to the energy coupling, and can be used to determine the position of the cantilever by scanning the cantilever in the xy-plane, see figure \ref{figure:figure1}d.

For the experiment presented in this paper, we positioned the cantilever above the center of the pickup coil, to minimize possible repulsive forces from the superconducting wires. By gently decreasing the height of the cantilever until the signal is completely lost, we determine the relative height of the magnetic particle with respect to the surface. The sample holder is placed on a finestage, machined out of aluminum, which can be moved in all spatial directions by actuating laminated piezoelectric extension stacks. Using this, we can now have good control of the height up to the full range of the finestage of $2.3$ $\mu$m \footnote{The piezoelectric extension stacks are of type P-883.51 by Physik Instrumente GmbH and Co. KG. Germany. To determine the range of the finestage, we extrapolated data from reference \onlinecite{Taylor2006} for the actuator constant from $20$ K to $0$ K.}.

We swept the drive frequency at a drive amplitude small enough to avoid non-linear responses of the cantilever's motion, while measuring the SQUID signal. We measured with a sampling time of $2$ s every $0.02$ Hz. Fitting the data with a Lorentzian, we obtain $f_r$ and $Q=\frac{f_r}{FWHM}$. At each height, the temperature was varied from the lowest achievable temperature $\approx25$ mK, up to $500$ mK. Above $500$ mK, the aluminum shielding of the experiment starts to become non-superconducting. An example of the data with the Lorentzian fits at all used heights at $70$ mK is shown in figure \ref{figure:figure2}.

\begin{table}[bbb]
\centering
\caption{Obtained values for the spin density $\sigma$ and relaxation time $T_1$ for every height $z$ above the sample. See \ref{figure:figure3} for the individual fits figure. The bottom row shows the average value and the standard deviation.}
\label{tabel}
\begin{tabular}{ccc}
\hline
\hline
Height ($\mu$m) 	& 		spin density (nm$^{-2}$) 	& Relaxation time (ms)  \\ \hline
$0.08$            	&       $0.142$      		   		&         $0.42$        \\
$0.19$            	&   	$0.137$      		       	&         $0.52$        \\
$0.38$            	&      	$0.140$      	     		&         $0.48$        \\
$0.57$            	&      	$0.142$      	     		&         $0.42$        \\
$0.77$            	&      	$0.136$           			&         $0.38$        \\
$1.15$            	&      	$0.130$          			&         $0.32$        \\
$1.72$            	&      	$0.133$             		&         $0.28$        \\
$2.30$            	&  		$0.168$						&         $0.33$        \\ 
\hline
mean:		 		&      	$0.14\pm0.01$	   			&      $0.39\pm0.08$ \\       
\hline
\hline        
\end{tabular}
\end{table}

The results of our measurements described above are shown in figure \ref{figure:figure3} together with the fits according to equations \eqref{eq:1} and \eqref{eq:2}. At every height $z$ and temperature $T$ we calculate the value for $C$ according to equation \eqref{eq:3}. The quality factor far from the surface $Q_0=2.8\cdot10^4$. The stiffness $k_0=7.0\cdot10^{-5}$ Nm$^{-1}$ of the cantilever is calculated using $k_0=m_{eff}\left(2\pi f_0\right)^2$ with $f_0=3.0$ kHz and $m_{eff}=2.0\cdot10^{-13}$ kg. The effective mass $m_{eff}$ is calculated using the geometry of the cantilever and the magnetic particle. The magnetic particle is taken as a spherical dipole with magnetic moment $\bm{m}$. According to the model, the temperature at which the resonance frequency close to the sample has a maximum, is independent of $\sigma$ and $T_1$, but is dependent on the absolute value of $\bm{m}$ and the distance to the sample. We find $m=1.9\cdot10^{-11}$ JT$^{-1}$. From this we find an effective saturation magnetization of $1.15$ T for a sphere that is fully magnetic. Alternatively we can assume $\mu_0 M_{sat}=1.3$ T and an outer layer of $200$ nm which is magnetically dead. The magnetic moment of the dangling bonds \cite{Haneman1968} is equal to the Bohr magneton $\mu=9.274\cdot10^{-24}$JT$^{-1}$.

The solid lines in figure \ref{figure:figure3} are fits to the data according to equations \eqref{eq:1} and \eqref{eq:2} with $\sigma$, $T_1$ and $f_0$ as the only fitting parameters. All fitting parameters are separately fitted for each height, for both the frequency data and the quality factor data. $f_0$ is a temperature independent parameter different for each height, which we attribute to an unknown mechanism, since the coupling to the SQUID is too small of an effect at these distances and has a height dependence with opposite sign to the one observed. The results of the fits for $T_1$ and $\sigma$ can be found in table \ref{tabel}. We left $\sigma$ as fitting parameter for each height, to verify the correctness of our analysis, since this number should be the same for each height. We see that $T_1$ slightly increases when the magnetic particle approached the surface, as is also observed for bulk spins in electron spin resonance experiments \cite{Stipe2001a}. $T_1$ could depend on temperature, but by taking the ratio of equation \eqref{eq:1} with equation \eqref{eq:2} we extract $T_1$ for each measurement, and we find that $T_1$ is constant with temperature to within $20$\%. The average values of all individual fits are $\sigma=0.14\pm0.01$ spins per nm$^2$ and $T_1=0.39\pm0.08$ ms. The found value for $\sigma$ is similar to values measured using Electron Paramagnetic Resonance \cite{Haneman1968, Haneman1975}.

The dashed line in figure \ref{figure:figure3} is the frequency shift calculated with the commonly used expression $k_s= p\bm{\mu} \cdot \frac{d^2 \bm{B}}{dx^2}$, with $p=\tanh{\left(\frac{\mu B(\bm{r})}{k_BT}\right)}$. Important is that for this curve, the spin density is ten times smaller than we find with our analysis. 

The deviation of the data from the fit for low temperatures and small values for $z$ can be understood by considering that we do not have only spins at the surface. Electron spins inside the bulk will cause deviations to the fits, already when the density is in the order of $10^4$ spins per $\mu m^3$ which is less than 1 ppm of the silicon atoms. Considering the nuclear spins, the $4.7$\% natural abundance of the $^{29}$Si isotope can only account for less than $1$ percent deviation.

Note that in electron spin resonance studies with our MRFM setup, a value for $T_1$ in the order of seconds was reported \cite{Vinante2011}.  With our new analysis we believe that it is possible that the reported long lived frequency shifts could be caused by nuclei polarized by interactions with these electron spins, and that these electron spins were actually much shorter lived, as is reported for nitroxide-doped perdeuterated polystyrene films \cite{Chen2013}.

Our analysis suggests that the spin mediated dissipation is the main mechanism leading to a significant reduction in the quality factor of the cantilever. Previous work at higher temperatures \cite{Kuehn2006} reports dielectric fluctuations as the main non-contact dissipation mechanism. We do not see any evidence in our measurements for this. Possibly, the use of a laser in the setup to read out the cantilever causes extra charge fluctuations. Furthermore, we work at lower temperatures, where the large spin polarization enhances the magnetic dissipation and possibly reduce fluctuating charges. 

We calculated the magnetic dissipation for a magnetic imaging experiments at higher temperature and a different tip-sample geometry. The results can be found in table \ref{tabel2}. We used the experimental parameters for a state-of-the-art MRFM \cite{Degen2009}. In this apparatus, the bare non-magnetic cantilever is centered approximately $50$ nm above a magnetic particle on the substrate, which is assumed for simplicity to be a spherical particle with a radius of $100$ nm. This setup is equivalent to a magnetic dipole attached to the cantilever itself approaching a surface with the shape of the cantilever. The magnetic dipole and external field are oriented in the z-direction while the fundamental mode of the cantilever is in the x-direction. For the cantilever, we used the parameters of a recently developed diamond cantilever \cite{Tao2014} which is shown to have at low temperatures an intrinsic quality factor $Q_0=1.5\cdot10^6$, resonance frequency $f_0=32$ kHz and stiffness $k_0=6.7\cdot10^{-2}$ Nm$^{-1}$. A spin density $\sigma=0.14$ nm$^{-2}$ is used, which is found in this report to be the density for the silicon surface, but it is also close to the typical values found for diamond surfaces \cite{Rosskopf2014}. Only spins at the very end of the cantilever are considered since this surface contributes most to the dissipation, which is $0.66$ $\mu$m thick and $12$ $\mu$m wide. Although equation \eqref{eq:1} cannot be used since we do not have a uniform infinite surface anymore, according to the original expressions \cite{DeVoogd2015} one can continue to use equation \eqref{eq:2} for the dissipation replacing the integral in equation \eqref{eq:3} over the end of the cantilever. The relaxation time is chosen such that the dissipation is maximum: $T_1=\left(2 \pi f_0\right)^{-1}=5.0$ $\mu$s. 

\begin{table}[tb]
\centering
\caption{Calculated quality factor Q for three different temperatures and external magnetic fields assuming magnetic dissipation as the only source for non-contact friction. Calculations are based on a state-of-the art MRFM apparatus with a `sample on cantilever' geometry \cite{Degen2009} and a cantilever \cite{Tao2014} with an internal quality factor $Q_0=1.5\cdot10^6$.}
\label{tabel2}
\textbf{Calculated quality factors ($\cdot10^6$)}
\begin{tabular}{m{2cm} m{2cm} m{2cm} m{2cm} }
\hline
\hline
   & $T=10$ mK & $T=300$ mK & $T=4.2$ K \\ \hline
 $B_{ext}=0$ T    & 0.49  & 0.20   & 0.98  \\ 
 $B_{ext}=0.1$ T  & 1.50  & 0.19   & 0.91  \\ 
 $B_{ext}=6$ T    & 1.50  & 1.50   & 1.17  \\
\hline
\hline
\end{tabular}
\end{table}

The values in table \ref{tabel2} show that the magnetic dissipation could be an important source of non-contact friction. Furthermore we see that applying external fields can reduce the magnetic dissipation. Considering these calculations we believe that the magnetic dissipation we find at low temperatures can be avoided with the correct choice for the substrate and the use of large external magnetic fields.  

To summarize, we have shown how the dissipation and frequency shift  mediated by spins in magnetic force experiments can be fully understood. The new analysis suggest that in order to achieve higher sensitivity in magnetic imaging techniques, one should not only focus on improving the intrinsic losses of the micro-mechanical cantilever, but also on the reduction of electron spins in the sample. Furthermore we have shown how the spin's relaxation time can be extracted without the use of resonance techniques. For silicon substrates with native oxides, we find a relaxation time of $T_1=0.39\pm0.08$ ms and a spin density of $\sigma=0.14\pm0.01$ per nm$^2$.  The understanding of the spin mediated dissipation is important to further improve the mechanical resonators in magnetic imaging experiments.  
 
\begin{acknowledgments}
We thank D. J. Thoen, A. Endo and T. M. Klapwijk for fabricating the NbTiN detection chips, M. Azarkh for magnetizing the magnetic particle, A. Singh for support with sample fabrication and F. Schenkel, J. P. Koning and D. J. van der Zalm for technical support. We thank L. Bossoni, B. van Waarde and M. de Wit for proofreading the manuscript. This work was supported by the Dutch Foundation for Fundamental Research on Matter (FOM), and by a VICI fellowship by the Netherlands Organization for Scientific Research (NWO) to T.H.O.

A. M. J. den Haan and J. J. T. Wagenaar contributed equally.
\end{acknowledgments}

\bibliography{bibliography}

\end{document}